\documentstyle[aps,prd,twocolumn]{revtex}
\input epsf
\begin{document}
\draft
\title{Quark flavor tagging in polarized hadronic processes}
\author{J.~M.~Niczyporuk\thanks{Present Address:  Department of Physics,
  University of Illinois, Urbana, Illinois  61801.} 
	 and E.~E.~W.~Bruins\thanks{Present Address:  Stichting FOM, Utrecht,
 		The Netherlands.}}
\address{Laboratory for Nuclear Science and Department of Physics,
  		Massachusetts Institute of Technology,\\
        Cambridge, Massachusetts  02139}
\date{Received 13 April 1998}
\maketitle

\begin{abstract}
We describe a general approach to quark flavor tagging in
polarized hadronic processes, with particular emphasis on 
semi-inclusive
deep inelastic scattering.
A formalism is introduced
that allows one to relate chosen quark flavor polarizations to an 
arbitrary combination of final-state hadron spin asymmetries.  
Within the context of the presented formalism, we 
quantify the sensitivity of various semi-inclusive hadron
asymmetries to the light quark flavors.  
We show that {\it unpolarized} $\Lambda$'s may allow one to
measure strange quark and antiquark polarizations independently.
We also highlight several applications of our formalism,
particularly to measurements intended to probe further the
spin structure of the nucleon.
\end{abstract}


\section{Introduction}
\label{section:intro}
The interpretation of hard hadronic processes in terms of partonic (quark
and gluon) degrees of freedom forms a vital component of 
modern research in particle physics.
In particular, the tagging of parton quantum numbers, such as flavor
and charge, using appropriately chosen final states allows important tests
of both QCD and electroweak dynamics.  
In the remainder of this paper, we will focus primarily on 
the leptoproduction of hadrons in deep inelastic scattering (DIS), 
$\ell N \rightarrow \ell' h X$, where both initial states are polarized.
Such measurements 
continue to provide
important information on the spin 
structure of the nucleon~\cite{spin_review}.
However, the conceptual foundations of our approach and the 
formalism that we introduce
pertain to any hard hadronic process, and other applications
are mentioned in Sec.~\ref{sec:applications}.  

The spin structure of the nucleon has received much attention in the past
decade~\cite{spin_review}.  
The experimental focus until rather recently has been
on the leading twist structure function $g_1(x,Q^2)$, which is roughly 
proportional
to the inclusive spin asymmetry on a longitudinally polarized target.
However, processes where at least one hadron is detected in the final state
offer several distinct advantages over the inclusive process 
alone~\cite{heinmann,feynman_close,frankfurt89,close91}.
In particular, semi-inclusive reactions provide a direct probe of the flavor  
dependence of quark observables, 
allowing more stringent tests of hadron structure.  
 Perhaps more importantly, semi-inclusive processes allow one to
 separate~\cite{frankfurt89,close91} contributions with 
 definite charge conjugation symmetry.  This is particularly interesting
 since charge-odd quantities are free of the axial anomaly.
 It has been recently suggested~\cite{mythesis}
 that a comparison of the charge conjugation even and odd
 parts of parton helicity and transversity 
 distributions may yield
 new information on the polarized gluon
 contribution to the nucleon spin.

Given this importance of semi-inclusive measurements in hard hadronic
processes, it is useful to have a formalism that is
independent of model assumptions. 
In Sec.~\ref{sec:formalism}, we introduce a general method to extract chosen
polarized quark distributions from an arbitrary combination of 
final state asymmetries.  This new approach is particularly
useful in light of the latest generation of existing [HERMES, Spin Muon
Collaboration (SMC), TJNAF] and
forthcoming [COMPASS, ELFE, HERA-$\vec{p}$, BNL Relativistic Heavy
Ion Collider (RHIC) Spin] experimental efforts,
which allow identification of a large number of final state hadrons. 
Within the framework of the presented formalism, we quantify the sensitivity, 
as a function of kinematics, of various hadrons to the light quark flavors.

\section{General Considerations}
\label{sec:general}
Traditionally, quark flavor tagging has been applied almost exclusively to 
the current fragmentation region in deep inelastic processes.
Since a separation between the current and target
regions is usually regarded as a necessary criterion for such
measurements, one imposes kinematic cuts in order to try to
exclude the target region.  Our point is that such a separation
is not possible, even in principle, within a coherent and complete description
of hadronic final states.\footnote{Sophisticated hadronization models,
such as the string and cluster models, reflect this fact.}
In perturbation theory,
this follows from the existence of collinear singularities with respect to
initial state partons that generate a contribution to the target fragmentation
region~\cite{graudenz94}.
We further argue that, generally speaking,
there is no {\it a priori}
reason to minimize the effects of target fragmentation.  
Indeed, it has been shown~\cite{graudenz94} that, independent of a
transverse momentum cutoff, all collinear singularities for
sufficiently hard hadrons can be absorbed into so-called fracture 
functions~\cite{trentadue94} $M_{i,N}^h(x,x_F,Q^2)$, 
which give the joint probability
of ``finding" within the target $N$ a parton $i$ and hadron $h$.
Measurements sensitive to target fragmentation 
offer complementary information on the nucleon state~\cite{target1}
and additional insights into a 
unified view~\cite{target2} of hadronic reactions.


Quark flavor tagging is based on the idea of
local parton-hadron duality (LPHD)~\cite{lphd},
where it is assumed that the flow of quantum numbers at
the hadron level tends to follow the flow established at the parton level.
At leading order in perturbation theory (for infrared safe or
factorizable parton quantities), this has the practical
consequence that the parton-hadron correspondence is essentially
one-to-one.
Since quarks and gluons are not asymptotic 
states (due to confinement), 
it would be a mistake to conclude that we can ever measure, in effect,
a primary parton, instead of merely an event property that is
{\it correlated} with the primary parton.      
Within a complete description of hadronic processes, valid beyond
the leading order, the relevant issue is the degree of correlation 
between the final state topology 
and the quantum numbers of the primary partons.  

It is therefore instructive to discuss briefly the nature of parton-hadron
correlations.
Inclusive hadron distributions in all hadronic processes,
regardless of whether the collisions are hard or soft,
are characterized by projectile fragmentation 
regions separated by
a central region~\cite{bjorken73}.  
The key feature of the central region is that it
is essentially independent of both projectiles and hence 
universal for all hadronic processes (at fixed invariant mass $W$).
This has been confirmed in hadron-hadron 
collisions~\cite{adamus88}, and indeed,
hadron spectra in the central region are found to be
very similar~\cite{aid95} 
in $pp$, $p\bar{p}$, photoproduction, and low-$x$ DIS processes.
These generic features of hadronic final states have been 
recognized~\cite{feynman69,bjorken73} a long time ago 
to be a consequence of Lorentz invariance and 
short range correlations in rapidity.

Using an approach based on short range 
rapidity correlations and LPHD, we have calculated~\cite{mythesis} 
the correlations of light hadrons with respect to the current quark and
target remnant.  
Within our discussion, it is important to distinguish between 
forward ($x_F>0$) and backward ($x_F<0$) regions, which are {\it defined}
strictly by kinematics, from the current, target, and central
fragmentation regions, which are never distinct, but represent varying
degrees of correlation with the quark and remnant.  
As the correlations depend strongly on rapidity differences,
our definition of the fragmentation regions coincides with their 
classification, within perturbative QCD, in terms of collinear
and soft divergences.
This suggests that it is favorable to use hadronic variables 
such as $x_F$ or rapidity
over the usual energy fraction $z=E_h/E_{\gamma}$, since the latter variable
cannot distinguish between the central
and target regions (both dominate at low $z$).

In our calculations, we find that at small $|x_F|<0.1-0.2$ and 
{\it fixed} $Q^2$, both the
current and target correlations decrease as $W$ 
increases.\footnote{Several quantitative examples were given in
Ref.~\cite{berger87} on the basis of scaling 
violations~\cite{siegrist82} in inclusive
charged particle momentum spectra in $e^+e^-$ collisions. 
However, as discussed in Refs.~\cite{siegrist82,halzen84}, these scaling
violations are actually due to threshold production of charmed particles,
where, as expected, the region of scaling extends down to lower values
of $x_F$ as $W$ is increased.}
This implies that most of the increase in hadron production
with increasing $W$ occurs in the central region, a result well
known from hadronic phenomenology.  Indeed, in DIS at {\it constant} $Q^2$,
the charged pion fragmentation functions $D^{\pi}(x_F,W)$ 
rise linearly with $\ln{W}$ at small $|x_F|<0.1$, but remain
constant at larger values of $|x_F|$, both in the forward 
and backward regions~\cite{arneodo86}.
This is an important result, 
because it illustrates that independent of the target remnant, 
there are kinematical effects in the forward
region that cannot be described by fragmentation functions depending
on $z$ and $Q^2$ alone, as is usually assumed.  
For increasing
values of $W$, while the current and target regions become better
separated kinematically, the fraction of hadrons that are strongly
correlated with the current quark decreases (roughly as $1/\ln W$).
Even in the central region, however, the correlation between an arbitrary
final state and the flavor of the current quark remains significant due to the
finite number of accessible flavors.
In the next section, we introduce a new formalism that allows one to
quantify the above correlations and to
relate chosen hadron-level asymmetries
to the polarization of partons involved in the hard scattering
process.

\section{Formalism}
\label{sec:formalism}
Hadronic final states in deep inelastic processes are typically analyzed in
terms of the usual fragmentation functions $D_i^h$~\cite{feynman_close}, 
which parameterize the probability
density for producing hadrons $h$, 
given an initial state parton of 
flavor $i$ ($i=u,\bar{u},d,\bar{d},\ldots$).  
The corresponding
formulas for polarized and unpolarized scattering, at leading twist,
are well known.  
However, fragmentation functions are not the most useful quantities in the
context of quark flavor tagging.  Instead, we define the    
{\it quark flavor purity} $P_i^h$ as the probability
that a quark of
flavor $i$ was probed by the virtual photon, 
given a final state $h$.   In terms of
the usual quantities, we can write
\begin{eqnarray}
		P_i^h & = & \frac{e_i^2 \; q_i \; D_i^h}
						{\sum_{i'=f,\bar{f}}\; e_{i'}^2\;q_{i'}\;D_{i'}^h}    
 \label{eq:purity}
\end{eqnarray}
where $e_i$ are the quark charges, $q_i=q_i(x,Q^2)$ are helicity averaged
quark distributions, 
$D_i^h=D_i^h(x_F,Q^2,x)$ are generalized
fragmentation functions,\footnote{For convenience, 
we integrate over the hadron's $p_T$ and $\phi$.} and
$\sum_{i=f,\bar{f}} \;P_i^h  =  1$ holds for any final state.
The purities not only have the physically meaningful interpretation mentioned
above, but also are the relevant quantities for performing a 
polarized quark flavor decomposition from semi-inclusive asymmetries. 

As a specific example, we consider semi-inclusive deep inelastic scattering
with both beam and target longitudinally polarized, where the photon
is the exchanged boson.
In this case, for hadrons $h$ in the final state,
one measures the virtual photon-nucleon asymmetry
\begin{eqnarray}
	A^h(x,x_F,Q^2) & = & 
      \frac{ \sigma^h_{1/2} - \sigma^h_{3/2} }
	       { \sigma^h_{1/2} + \sigma^h_{3/2} }	  
\label{eq:asymmetry_def}
\end{eqnarray}
where $\sigma^h_{1/2\,(3/2)}(x,x_F,Q^2)$ denotes
the cross section when the projection of the total angular momentum
of the photon-nucleon system is $1/2(3/2)$.
We can then write any such asymmetry (at leading twist)
in terms of only purities and
quark flavor polarizations $(\Delta q/q)_i$:
\begin{eqnarray}  
	 A^h & = & \sum_{i=f,\bar{f}} P_i^h \; 
			{\left({ {\Delta q} \over q }\right)_i} .
\label{eq:asymmetry}
\end{eqnarray}
Following the arguments in the previous section, Eq.\ (\ref{eq:asymmetry}) 
holds in both the forward and backward regions, for sufficiently
hard hadrons.   
The derivation of Eq.\ (\ref{eq:asymmetry})  depends crucially 
on the independence of the fragmentation process on the quark
polarization.  Although this is usually stated as an additional assumption,
it is actually a consequence of 
the parity invariance of the strong interaction:
for inclusive production of unpolarized hadrons, there is no pseudovector
observable in the final state that could in principle couple to the
quark polarization.
It follows that, for example, the presence of target fragmentation 
does not dilute the asymmetry.  In general, there is no reason to 
minimize the
contribution of target fragmentation, as long as the correct
purities are used in the analysis.    

The form of Eq.\ (\ref{eq:asymmetry}) allows us to combine several
independent asymmetries  into a column vector, yielding the matrix
equation:
\begin{eqnarray}
\vec{A} & = &
  {\cal P} \stackrel{\longrightarrow}{\left( \frac{\Delta q}{q} \right)} .
\label{eq:puritymatrix}
\end{eqnarray}
When the matrix elements of ${\cal P}$ are determined (from data or
simulations), Eq.\ (\ref{eq:puritymatrix}) contains {\it all} information 
needed to extract specified polarized quark distributions from a
chosen set of semi-inclusive asymmetries.  
The inclusive asymmetry, where only the scattered lepton is detected,
also contains useful information and can be trivially incorporated
within the purity formalism:
\begin{eqnarray}
		P_i^{incl} & = & \frac{e_i^2 \; q_i}
			{\sum_{i'=f,\bar{f}} \; e_{i'}^2\;q_{i'}}     
 \label{eq:inclusive_purity}
\end{eqnarray}	
Within the presented 
approach, then, the extraction
of polarized quark flavor distributions is reduced to optimization,
via choice of final (and initial) state hadrons and kinematics, of the purity 
matrix.  Several methods of optimization 
are suggested in Ref.~\cite{mythesis}.  

We have studied the kinematical dependence of purities for light
quark flavors and hadrons, on proton and deuteron targets,
using {\sc lepto}-6.1~\cite{lepto} for event generation 
and {\sc jetset}-7.4~\cite{jetset} 
for hadronization.  Since the purities are independent of
the quark helicities, an unpolarized simulation can be used to
generate them.
We have used the HERMES kinematics
and a parametric model~\cite{holger} of its acceptance in our 
simulations (see Ref.~\cite{mythesis} for details), applying standard
DIS selection cuts of $Q^2>1$ GeV$^2$, $W^2>4$ GeV$^2$, and $y<0.85$.
The following numerical results therefore apply 
primarily to the forward region,
as HERMES detects mostly forward hadrons.
As expected from
the form of Eq.\ (\ref{eq:purity}), the purities are relatively 
insensitive to the choice of 
hadronization model 
parameters and unpolarized parton distributions.      
\begin{figure}[t!]
\epsfxsize=8.6cm \epsfbox{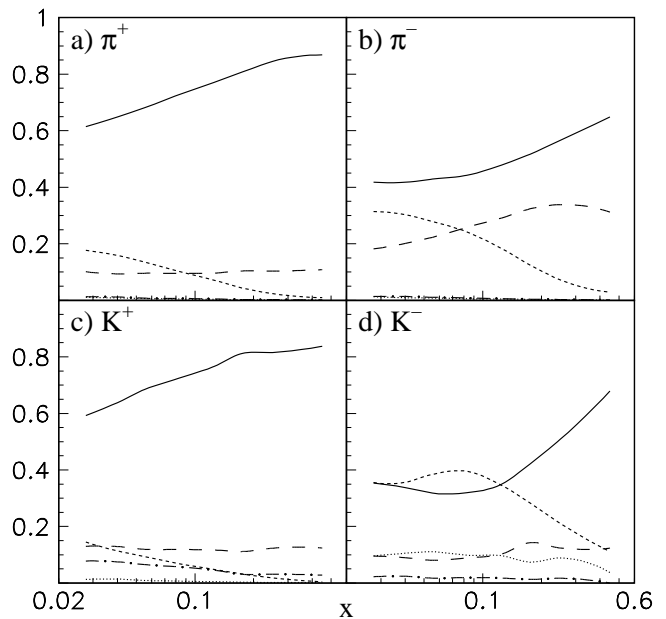} 
\caption{Purities $P_f^h(x,|x_F|>0.1)$ for typical final states with 
respect to the light quark flavors: $u$ (solid), 
$\bar{u}$ (short-dashed), $d$ (long-dashed), $s$ (dotted),
and $\bar{s}$ (dot-dashed).  The $\bar{d}$-quark contribution never
exceeds 10\% and is not shown.}
\label{fig1}
\end{figure}

In Fig.\ \ref{fig1}, we plot quark flavor purities for typical final
states as a function of the Bjorken-$x$ variable, using a deuteron target
(results on the proton are qualitatively similar).  All quark flavor 
labels refer to the proton, using isospin symmetry.  The cut $|x_F|>0.1$
is applied to suppress very central hadrons.  
Using the purities shown, it is straightforward to quantify the
sensitivity of semi-inclusive asymmetries to
the light quark and antiquark flavors.  We summarize our general
conclusions here.  As expected from charge counting,  
positively charged hadrons are primarily sensitive
to the $u$-quark [see Figs.\ \ref{fig1}(a),(c)].
Given that the inclusive asymmetry on a proton target
essentially probes the combination 
$\Delta u + \Delta \bar{u}$, one can reasonably constrain the 
$u$- and $\bar{u}$-quark polarizations from a combined analysis
of inclusive and positively charged hadron data alone.
Since $P_f^{\pi^-}$ is dominated by $u$-, $\bar{u}$-, and $d$-quarks at 
small $x<0.1$ [see Fig.\ \ref{fig1}(b)], including $A^{\pi^-}$ in the 
above analysis allows one to extract $\Delta u/u$, $\Delta \bar{u}/\bar{u}$, 
and $\Delta d/d$ in the sea region.  In addition, as only
$u$- and $d$-quarks contribute significantly to $P_f^{\pi^-}$ in the
valence region ($x>0.2$), one can extract $\Delta u/u$ and
$\Delta d/d$ (at large $x$) using {\it only} $A^{\pi^+}$ and $A^{\pi^-}$. 
We conclude that a combination of asymmetries using 
copiously produced hadrons can
be used to extract $\Delta u/u$,
$\Delta \bar{u}/\bar{u}$, and $\Delta d/d$.
On the other hand, it will be a 
challenge to 
measure the polarization of $\bar{d}$-quarks, since the corresponding purities
never exceed 10\% for the final states shown in Fig.\ \ref{fig1}.  
We suggest combining $\Delta d$ with a measurement of
$\Delta d_v$ ($\equiv \Delta d - \Delta \bar{d}$) using
the pion charge difference asymmetry method of 
Ref.\ \cite{frankfurt89}.\footnote{The derivation in Ref.\ \cite{frankfurt89} 
assumes charge and isospin conjugation symmetries for the
fragmentation functions, which are formally violated in a 
complete QCD description of DIS.  Phenomenological consequences
are studied in Ref.\ \cite{deflorian96}.}

\begin{figure}[t!]
\epsfxsize=8.6cm \epsfbox{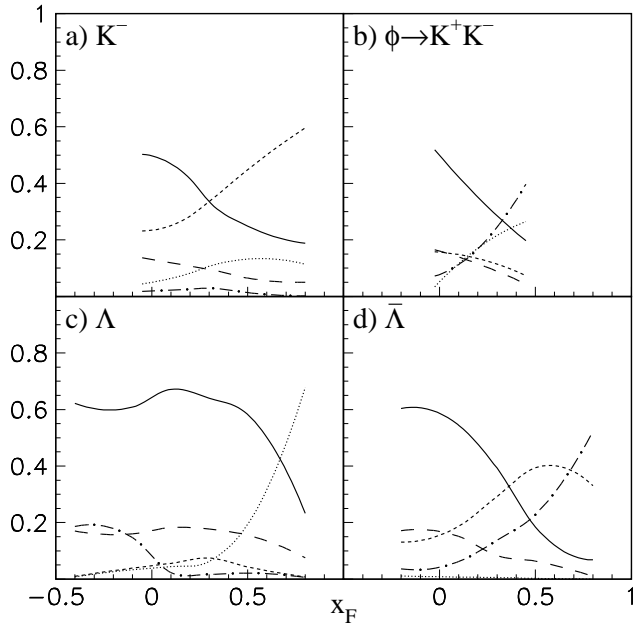} 
\caption{Purities $P_f^h(x_F,x<0.1)$ for strange hadrons with
respect to the light quark flavors: $u$ (solid), 
$\bar{u}$ (short-dashed), $d$ (long-dashed), $s$ (dotted),
and $\bar{s}$ (dot-dashed).  The $\bar{d}$-quark contribution never
exceeds 10\% and is not shown.  In (b), the $x_F$ variable refers to the
daughter $K^+$.}
\label{fig2}
\end{figure}

We have also studied the sensitivity of semi-inclusive
asymmetries to polarized strange quarks and antiquarks in the nucleon.
In Fig.\ \ref{fig2}, we plot the purities of selected strange hadrons 
as a function of $x_F$ in the sea region ($x<0.1$).    
The $K^-$ is often 
regarded as an interesting probe of the nucleon, as it is an 
``all sea" object.  We indeed find a large sensitivity to sea
quarks.  As expected from charge counting, the $\bar{u}$-quark dominates
over the $s$-quark contribution, independently of 
$x_F$ [see Fig.\ \ref{fig2}(a)].
On the other hand, selecting charged kaons from $\phi$ meson decay
provides improved sensitivity to
$s$- and $\bar{s}$-quarks [see Fig.\ \ref{fig2}(b)].
For example, using the cuts $x<0.1$ and $x_F>0.1$, we find that 
$P_s^{\phi}+P_{\bar{s}}^{\phi}\simeq40-45\%$.
The $\phi$ meson may
therefore be a useful probe of the polarized strange sea, provided one applies
kinematical cuts to suppress elastic production.  

Semi-inclusive $\Lambda$ ($\bar{\Lambda}$) asymmetries
are predominantly sensitive to $u$- ($u$- and $\bar{u}$-) quark
polarizations, 
except at rather large values of $x_F$ [see Figs.\ \ref{fig2}(c),(d)].  
However, we find that
the $\Lambda$ is sensitive, at the 20\% level, to $\bar{s}$-quarks at 
slightly negative $x_F<-0.2$, where the event rate is still 
reasonably large.  The key point is that the $\Lambda$ asymmetry 
in this region is
determined essentially by $\Delta u/u$, $\Delta d/d$, and 
$\Delta \bar{s}/\bar{s}$ (and the corresponding purities).  
Since $\Delta u/u$ and $\Delta d/d$ will be well
constrained by charged hadron and inclusive asymmetries (as discussed above), 
$\Lambda$ production in the backward region may be used to constrain the
polarized $\bar{s}$-quark sea.  In this case, to a good approximation, 
one can write
\begin{eqnarray}
	{ {\Delta \bar{s}} \over {\bar{s}}} \simeq
   	{1 \over {P_{\bar{s}}^{\Lambda}}} \; 
	{\left[A^{\Lambda} - \sum_{i=u,d} \; P_i^{\Lambda} \;
                        {\left({ {\Delta q} \over q }\right)_i}\right]} .
\label{eq:sbar}
\end{eqnarray}
Likewise, as the $\Lambda$ event rate distribution in $x_F$ is rather broad, 
one obtains appreciable sensitivity
to polarized $s$-quarks using only a modest cut on $x_F$.  
In particular, for $x_F>0.25$
(and $x<0.1$), we find that $P_u^{\Lambda}\simeq55\%$ and
$P_s^{\Lambda}\simeq P_d^{\Lambda}\simeq 20\%$.
Hence, given the contribution of $u$- and $d$-quarks, 
the ${\Lambda}$ asymmetry in this region probes $\Delta s/s$.
We therefore emphasize the importance of measuring $\Lambda$ asymmetries
in {\it both} the forward and backward regions.  

\section{Other Applications}
\label{sec:applications}
Thus far, we have introduced the quark flavor purities with 
particular emphasis
on their key role in interpreting electroproduction of unpolarized hadrons, 
where both beam and
target are longitudinally polarized.  
However, it is clear that our 
considerations generalize directly to all sufficiently hard hadronic
processes.  In general, the purities allow us to quantify,
in a well defined way, the sensitivity of chosen hadrons to the 
flavor of quarks involved in the
hard scattering process.  For the special case of 
polarization in the initial state,
the purities  relate, within the framework of a compact
formalism, hadron-level asymmetries to the 
polarization of initial state partons.  For example, one may imagine 
using the
purity formalism to measure the polarization of various quark flavors
using identified hadrons
in electroweak interactions, where parity violation provides
polarization observables for free.  
We leave these possibilities to
future work. 
We will focus instead on one example of particular
interest in spin physics:  deep inelastic production of polarized hadrons
(or jets).      

It is well known that polarized quarks can give rise to parity-odd 
correlations in the final state~\cite{collins95}.  
We illustrate here how virtual photoproduction of vector polarized final 
states (see Ref.~\cite{lambda}) 
can be conveniently analyzed using a formalism based on purities.
Our results, written explicitly for virtual photon-nucleon scattering, 
generalize trivially for other hard hadronic processes.
For definiteness, we choose a helicity basis oriented along the virtual
photon direction
($+\hat{z}$) and neglect finite angle effects, so that the fragmentation 
helicity basis is collinear with the quark 
helicity basis.  We then combine the polarizations $\rho^h\hat{z}$ of
chosen final states $h$ (possibly specified by kinematics and choice
of target) into a column vector $\vec{\rho}=(\{\rho^h\})$.
Our result is that at leading twist, for an {\it unpolarized} beam
incident on a target with longitudinal polarization $+\hat{z}$, 
\begin{eqnarray}
\vec{\rho} & = & 
 \Delta {\cal P} \stackrel{\longrightarrow}{\left( \frac{\Delta q}{q} \right)}
\label{eq:lambda_tarpol}
\end{eqnarray}  
where the matrix elements of $\Delta{\cal P}$ are defined by
\begin{eqnarray}
  \Delta P_i^h & = & P_i^h \; \left(\frac{\Delta D}{D}\right)_i^h .
\label{eq:spinpurity}
\end{eqnarray}
The $\Delta D_i^h$ are helicity difference fragmentation 
functions, which are assumed to be related, modulo dilution factors,
to the quark helicity structure 
of the final state.
Also, just as the purities satisfy the constraint
$\sum_{i=f,\bar{f}} \; P_i^h  =  1$, the 
quantities $\Delta P_i^h$ that enter Eq.\ (\ref{eq:lambda_tarpol})
are normalized by the longitudinal polarizations $\rho^h\hat{z}$ 
using a {\it polarized} beam ($+\hat{z}$) and an {\it unpolarized} target: 
\begin{eqnarray}
        \rho^h & = & \sum_{i=f,\bar{f}} \Delta P_i^h .
\label{eq:lambda_tarunpol}
\end{eqnarray}
Equations (\ref{eq:lambda_tarpol})-(\ref{eq:spinpurity}) also hold for
transversely polarized hadrons produced off a transversely polarized
target, provided we everywhere make the replacement
$\Delta \rightarrow \Delta_{\perp}$ (helicity$\rightarrow$transversity).
Indeed, formulas such as 
Eqs.\ (\ref{eq:lambda_tarpol})-(\ref{eq:lambda_tarunpol}) are generic
to processes with polarization observables
in the final state, where the $(\Delta D /D)_i^h$ are
the corresponding analyzing powers.      
We therefore conclude that the purities previously defined are also the
natural quantities for performing a global analysis of polarized hadron
(or jet) production in hard processes.

\section*{Acknowledgments}
 
We have benefited from discussions with R.~J.~Holt,
R.~L.~Jaffe, X.~Ji, L.~Mankiewicz, and R.~P.~Redwine.
This work is supported in part through
funds provided by the U.S. Department of Energy under Contract No.
DE-FC02-94ER40818.


%
%

%
%

\end{document}